\documentclass[%
reprint,
superscriptaddress,
nofootinbib,
 amsmath,amssymb,
 aps,
floatfix,
]{revtex4-2}

\usepackage{graphicx}
\usepackage{setspace}
\usepackage{dcolumn}
\usepackage{bm}
\usepackage{hyperref}
\usepackage[mathlines]{lineno}
\usepackage{amsmath, amssymb}
\usepackage{hyperref}
\usepackage{color}
\usepackage{rotating}
\usepackage{natbib}
\usepackage{orcidlink}

\begin{document}

\preprint{Draft}

\title{First Measurement of $a^0_2(1320)$ Polarized Photoproduction Cross Section}

\affiliation{Polytechnic Sciences and Mathematics, School of Applied Sciences and Arts, Arizona State University, Tempe, Arizona 85287, USA}
\affiliation{Department of Physics, National and Kapodistrian University of Athens, 15771 Athens, Greece}
\affiliation{Ruhr-Universit\"{a}t-Bochum, Institut f\"{u}r Experimentalphysik, D-44801 Bochum, Germany}
\affiliation{Helmholtz-Institut f\"{u}r Strahlen- und Kernphysik Universit\"{a}t Bonn, D-53115 Bonn, Germany}
\affiliation{Department of Physics, Carnegie Mellon University, Pittsburgh, Pennsylvania 15213, USA}
\affiliation{Department of Physics, The Catholic University of America, Washington, D.C. 20064, USA}
\affiliation{Department of Physics, University of Connecticut, Storrs, Connecticut 06269, USA}
\affiliation{Department of Physics, Duke University, Durham, North Carolina 27708, USA}
\affiliation{Department of Physics, Florida International University, Miami, Florida 33199, USA}
\affiliation{Department of Physics, Florida State University, Tallahassee, Florida 32306, USA}
\affiliation{Department of Physics, The George Washington University, Washington, D.C. 20052, USA}
\affiliation{School of Physics and Astronomy, University of Glasgow, Glasgow G12 8QQ, United Kingdom}
\affiliation{GSI Helmholtzzentrum f\"{u}r Schwerionenforschung GmbH, D-64291 Darmstadt, Germany}
\affiliation{Institute of High Energy Physics, Beijing 100049, People's Republic of China}
\affiliation{Department of Physics, Indiana University, Bloomington, Indiana 47405, USA}
\affiliation{National Research Centre Kurchatov Institute, Moscow 123182, Russia}
\affiliation{Department of Physics, Lamar University, Beaumont, Texas 77710, USA}
\affiliation{Department of Physics, University of Massachusetts, Amherst, Massachusetts 01003, USA}
\affiliation{National Research Nuclear University Moscow Engineering Physics Institute, Moscow 115409, Russia}
\affiliation{Department of Physics, Mount Allison University, Sackville, New Brunswick E4L 1E6, Canada}
\affiliation{Department of Physics, Norfolk State University, Norfolk, Virginia 23504, USA}
\affiliation{Department of Physics, North Carolina A\&T State University, Greensboro, North Carolina 27411, USA}
\affiliation{Department of Physics and Physical Oceanography, University of North Carolina at Wilmington, Wilmington, North Carolina 28403, USA}
\affiliation{Department of Physics, Old Dominion University, Norfolk, Virginia 23529, USA}
\affiliation{Department of Physics, University of Regina, Regina, Saskatchewan S4S 0A2, Canada}
\affiliation{Departamento de Física, Universidad T\'ecnica Federico Santa Mar\'ia, Casilla 110-V Valpara\'iso, Chile}
\affiliation{Department of Mathematics, Physics, and Computer Science, Springfield College, Springfield, Massachusetts, 01109, USA}
\affiliation{Thomas Jefferson National Accelerator Facility, Newport News, Virginia 23606, USA}
\affiliation{Laboratory of Particle Physics, Tomsk Polytechnic University, 634050 Tomsk, Russia}
\affiliation{Department of Physics, Tomsk State University, 634050 Tomsk, Russia}
\affiliation{Department of Physics and Astronomy, Union College, Schenectady, New York 12308, USA}
\affiliation{Department of Physics, Virginia Tech, Blacksburg, VA 24061, USA}
\affiliation{Department of Physics, Washington \& Jefferson College, Washington, Pennsylvania 15301, USA}
\affiliation{Department of Physics, William \& Mary, Williamsburg, Virginia 23185, USA}
\affiliation{School of Physics and Technology, Wuhan University, Wuhan, Hubei 430072, People's Republic of China}
\affiliation{A. I. Alikhanyan National Science Laboratory (Yerevan Physics Institute), 0036 Yerevan, Armenia}
\author{F.~Afzal\orcidlink{0000-0001-8063-6719 }} \affiliation{Helmholtz-Institut f\"{u}r Strahlen- und Kernphysik Universit\"{a}t Bonn, D-53115 Bonn, Germany}
\author{C.~S.~Akondi\orcidlink{0000-0001-6303-5217}} \affiliation{Department of Physics, Florida State University, Tallahassee, Florida 32306, USA}
\author{M.~Albrecht\orcidlink{0000-0001-6180-4297}} \affiliation{Thomas Jefferson National Accelerator Facility, Newport News, Virginia 23606, USA}
\author{M.~Amaryan\orcidlink{0000-0002-5648-0256}} \affiliation{Department of Physics, Old Dominion University, Norfolk, Virginia 23529, USA}
\author{S.~Arrigo} \affiliation{Department of Physics, William \& Mary, Williamsburg, Virginia 23185, USA}
\author{V.~Arroyave} \affiliation{Department of Physics, Florida International University, Miami, Florida 33199, USA}
\author{A.~Asaturyan\orcidlink{0000-0002-8105-913X}} \affiliation{Thomas Jefferson National Accelerator Facility, Newport News, Virginia 23606, USA}
\author{A.~Austregesilo\orcidlink{0000-0002-9291-4429}} \affiliation{Thomas Jefferson National Accelerator Facility, Newport News, Virginia 23606, USA}
\author{Z.~Baldwin\orcidlink{0000-0002-8534-0922}} \affiliation{Department of Physics, Carnegie Mellon University, Pittsburgh, Pennsylvania 15213, USA}
\author{F.~Barbosa} \affiliation{Thomas Jefferson National Accelerator Facility, Newport News, Virginia 23606, USA}
% ...
\collaboration{The \textsc{GlueX} Collaboration}

\date{\today}

\begin{abstract}
We measure for the first time the differential photoproduction cross section $d\sigma/dt$ of the $a_2(1320)$ meson at an average photon beam energy of 8.5~GeV, using data with an integrated luminosity of 104~pb$^{-1}$ collected by the GlueX experiment.  
We fully reconstruct the $\gamma p \to \eta\pi^0 p$ reaction and perform a partial-wave analysis in the $a_2(1320)$ mass region with amplitudes that incorporate the linear polarization of the beam. This allows us to separate for the first time the contributions of natural- and unnatural-parity exchanges.
These measurements provide novel information about the photoproduction mechanism, which is critical for the search for spin-exotic states.
\end{abstract}

\maketitle

\noindent\textit{Introduction~---} Quantum Chromodynamics (QCD) is the theory of the strong interaction, and describes the interaction of quarks mediated by force-carrying gluons. Most of the observed states bound by the strong interaction can be classified as either $q\bar{q}$ mesons or $qqq$ baryons.  However, QCD allows for the existence of a richer variety of bound states, and evidence for the existence of such states has been growing in recent years~\cite{Karliner:2017qhf,Olsen:2017bmm,Guo:2017jvc,Lebed:2016hpi,Chen:2022asf,Brambilla:2019esw,JPAC:2021rxu}.  One class of these states is the hybrid mesons, in which the confining gluonic field is excited and directly contributes to the properties of the meson. 
Hybrid mesons have been studied in various phenomenological~\cite{Szczepaniak:2001rg,Szczepaniak:2006nx,Guo:2008yz,Bass:2018xmz} and lattice QCD calculations~\cite{Lacock:1996ny,MILC:1997usn,Dudek:2013yja}, and have been searched for by many experiments~\cite{Meyer:2015eta}.

The best experimental candidate for the lightest hybrid meson is the spin-exotic $\pi_1(1600)$~\cite{ParticleDataGroup:2024cfk}.  
The $\pi_1(1600)$ has been studied most often by pion-production experiments, and has been seen to decay into $\eta\pi$ and $\eta'\pi$. To understand the nature of this state better, independent observations in different production mechanisms are crucial.
Identification of the $\pi_1(1600)$ in photoproduction would be a stepping stone on the path toward establishing a spectrum of hybrid mesons.
We also note that recently BESIII has reported the observation of a possible isospin-0 partner of the $\pi_1$ in its decay to $\eta\eta'$~\cite{BESIII:2022riz,BESIII:2022iwi}.

The first step towards identifying the $\pi_1(1600)$ in photoproduced $\eta^{(}{'}\vphantom{\eta}^{)}\pi^0$ is to study the production of the well-known $a_2(1320)$. The $a_2(1320)$ has a branching fraction to $\eta\pi^0$ that is about $30$ times larger than that to $\eta'\pi^0$, which makes the $\eta\pi^0$ decay preferred for this study.  The $a_2(1320)$ is a dominant contribution to the $\eta\pi^0$ spectrum and produces a clear peaking structure. It can therefore serve as a reference state for a partial-wave analysis of this channel, where, for example, the variation of the phase difference as a function of invariant mass between the $P$- and $D$-waves would be a signature for the existence of spin-exotic resonances.  Finally, obtaining a consistent description of photoproduction of well-known isovector mesons, such as the $\pi$ and $a_2(1320)$, will guide our modeling of the exotic isovector $\pi_1(1600)$. 

Most of the previous measurements of $\gamma p \to \eta\pi^0 p$ were performed at beam energies $E_\gamma<2.5$~GeV~\cite{GRAAL:2008,CrystalBallatMAMI:2009lze,cbelse:2008,CBELSATAPS:2014wvh,mainz:2018}, where the reaction is dominated by baryon resonances.  The only measurement at higher beam energies, where peripheral meson production dominates, was recently performed by the CLAS Collaboration using an unpolarized photon beam with $3.5 < E_\gamma < 5.5$~GeV~\cite{CLAS:2020rdz}.  They identified contributions from the $a_0(980)$ and $a_2(1320)$ mesons, and measured the differential photoproduction cross section of the $a_2(1320)$ as a function of
squared four-momentum transfer $t$.
To describe the CLAS photoproduction data for the $a_2(1320)$ and the $f_2(1270)$, the Joint Physics Analysis Center (JPAC) developed a model based on Regge theory assuming dominant contributions from vector (natural-parity) and axial-vector (unnatural-parity) Regge exchanges using two different assumptions for the helicity couplings at the photon-Reggeon-meson vertex~\cite{Mathieu:2020zpm}.
Under both assumptions, the model is qualitatively consistent with similar models describing $\pi^0$ photoproduction~\cite{JointPhysicsAnalysisCenter:2017del,Mathieu:2018mjw}, but neither of the two satisfactorily describes all the CLAS data. To further improve the modeling of this reaction, additional measurements are desirable at higher beam energy where Regge models may be more applicable, and with a polarized photon beam that allows disentangling of the vector and axial-vector exchange contributions.

In this Letter, we report on the measurement of the differential cross section of $a_2(1320)$ photoproduction using a linearly polarized photon beam with $8.2 < E_\gamma < 8.8$~GeV, in the reaction
\begin{equation}
    \gamma p \to a_2^0(1320) p ,~~  a_2^0(1320) \to \eta\pi^0 ,~~ \eta, \pi^0\to\gamma\gamma
\end{equation}
using a data sample with an integrated luminosity of 104~pb$^{-1}$.
The differential cross section is measured by performing a partial-wave analysis in five bins of $t$, in the range $0.1 < -t < 1.0$~GeV$^2$. The contributions to these cross sections from positive and negative reflectivities are determined separately for the first time. 

\begin{figure*}[!tb]
    \includegraphics[width=0.4\textwidth]{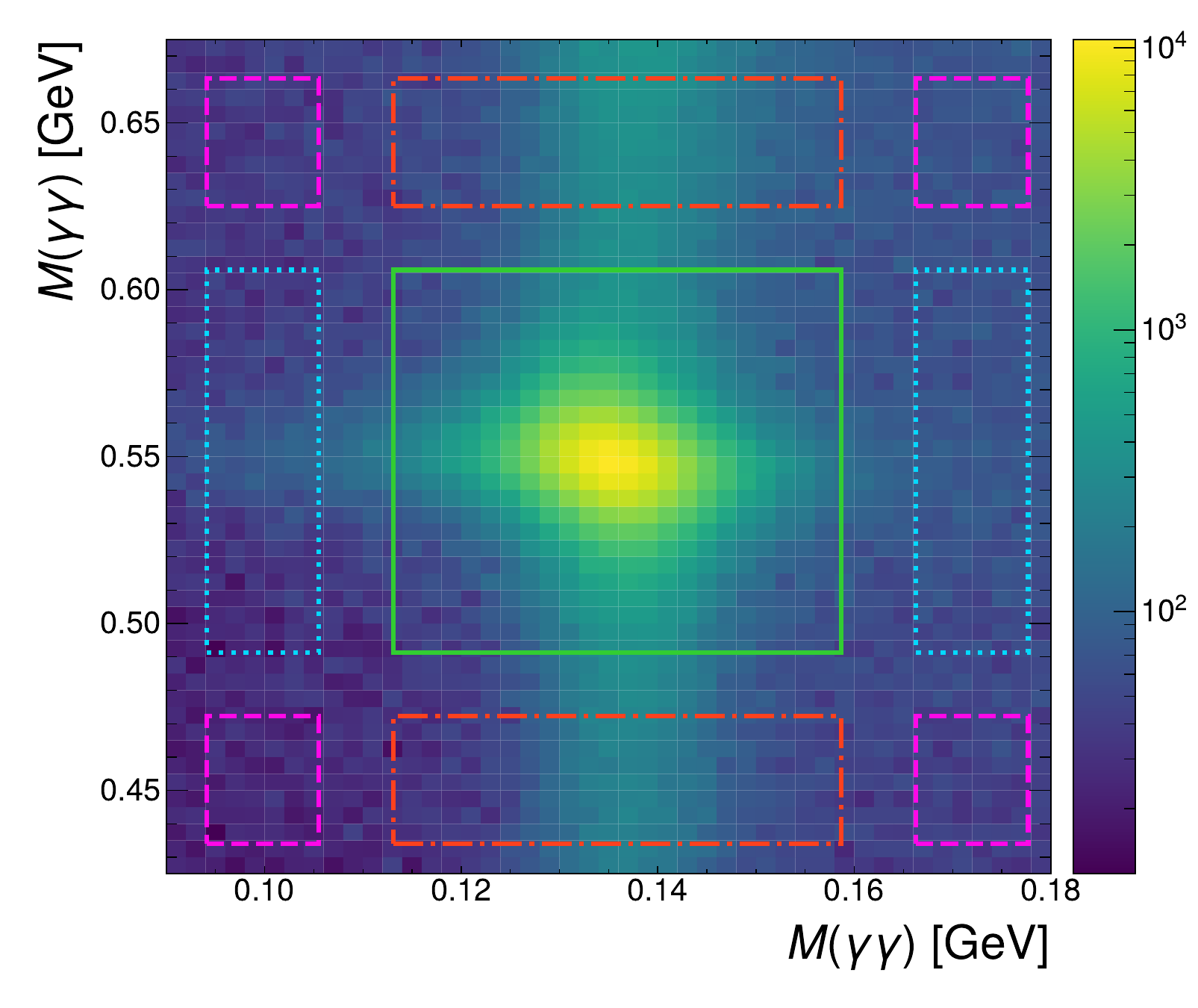}
    \includegraphics[width=0.4\textwidth]{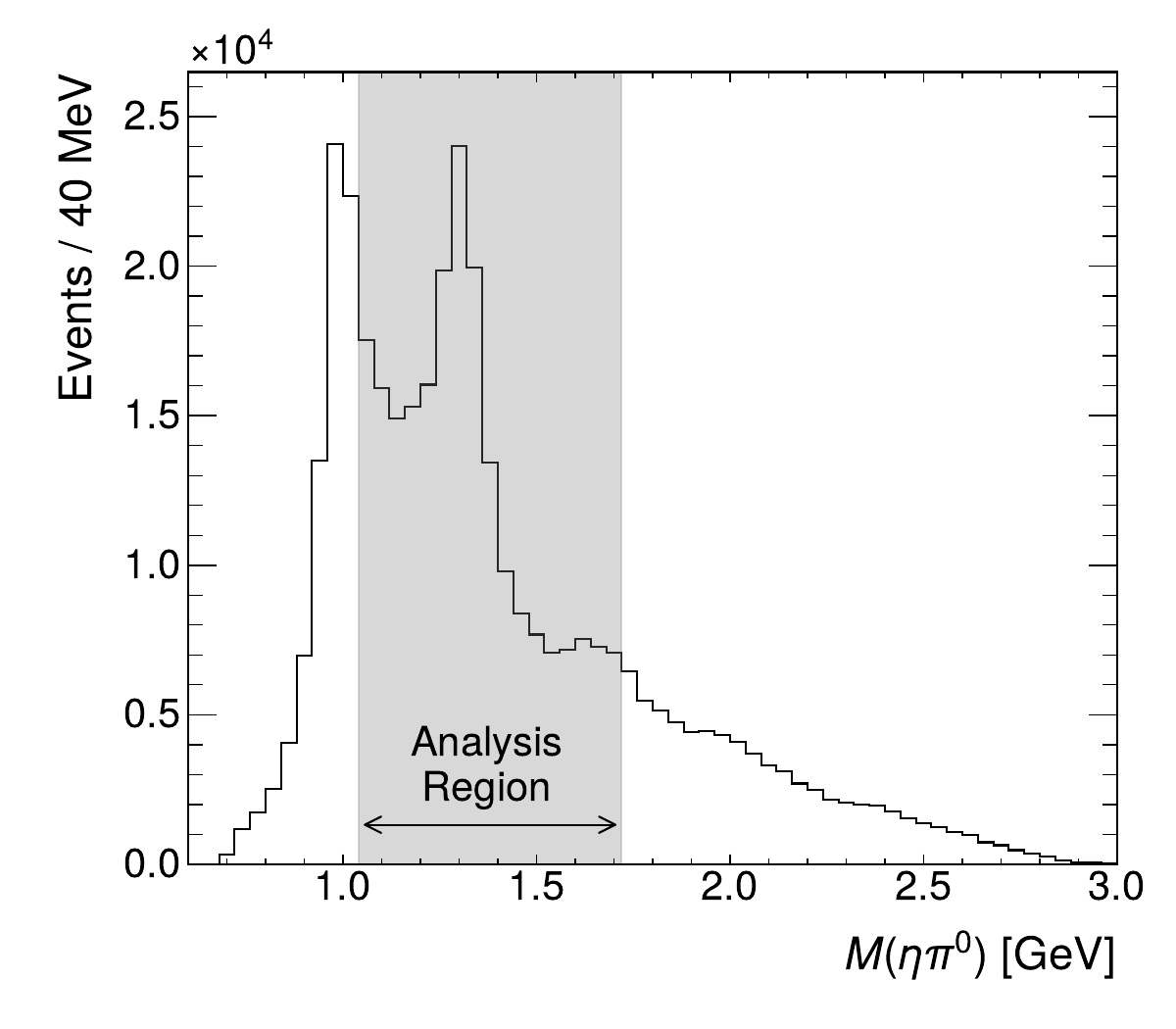}
    \caption{(Left) Two-dimensional distribution of the mass of $\eta$ candidates versus the mass of $\pi^0$ candidates.  The colored boxes indicate (green) the signal region, (red) the $\eta$ sideband, (cyan) the $\pi^0$ sideband, and (magenta) the corner sidebands. (Right) The invariant mass distribution for $\eta\pi^0$ candidate events after all event selections and background subtractions have been applied. Clear peaks around the nominal $a_0(980)$ and $a_2(1320)$ masses are observed. The shaded region around the $a_2(1320)$ indicates the mass range used for the partial-wave analysis described in the text.}
    \label{fig:4gmass}
\end{figure*}

\noindent\textit{Experimental Apparatus and Simulations~---}   The GlueX experiment consists of a tagged photon beam and a large-acceptance spectrometer, and has been described previously in detail in Refs.~\cite{GlueX:2020idb,Barbosa:2015bga,Berdnikov:2015jja,Dugger:2017zoq,Pentchev:2017omk,Beattie:2018xsk,Pooser:2019rhu,Jarvis:2019mgr}.  The photon beam is produced in bunches separated by 4~ns through coherent bremsstrahlung in a diamond radiator, with an energy spectrum that peaks at $E_\gamma= 8.8$~GeV and an average linear polarization fraction of $P\approx35\%$.  Approximately equal amounts of data are collected with polarization orientations of $-45^\circ$, $0^\circ$, $45^\circ$, and $90^\circ$ with respect to the floor of the experimental hall. 
The photon beam impinges on a 30~cm long liquid hydrogen target which sits inside of a 2~T solenoid.  
The target is surrounded by several drift chambers, calorimeters, and timing detectors which allow for the reconstruction and identification of charged and neutral particles over most of the solid angle.

The acceptance and reconstruction efficiencies are studied using events generated with the AmpTools package~\cite{amptools}, then passed through a Geant4-based Monte Carlo simulation of the detector~\cite{GEANT4:2002zbu}, and finally analyzed in the same manner as the measured data.  The simulated $\gamma p \to \eta \pi^0 p$ events used for acceptance corrections are generated to have the same $t$ distribution as observed for the measured data, but with isotropic angular distributions for the $\eta$ and $\pi^0$.

\noindent\textit{Event Selections~---}  We select events which have exactly four photon candidates and one recoil proton candidate. Beam photons are required to be in the energy range of $8.2 < E_\gamma < 8.8$~GeV, where the photon polarization is maximal. We select fully reconstructed events by performing a four-constraint kinematic fit imposing energy and momentum conservation. Additional details of these basic event selections which are common to most GlueX analyses are given in Sec.\,I of the Supplemental Material~\cite{SuppMaterials}
\nocite{VanHove:1969xa,Pauli:2018srg,Mathieu:2020zpm}.

We require $0.1 < -t < 1.0~\mathrm{GeV}^2$ to select reactions dominated by diffractive meson production.
Contributions from target-excitation reactions such as $\gamma p \to (\eta \Delta^*,\pi^0 N^*) \to \eta \pi^0 p$ are generally small and are rejected, as discussed in Sec.\,II of the Supplemental Material~\cite{SuppMaterials}.

The distribution of the two-photon invariant mass of one pair of photons versus the other pair of photons is shown in Fig.~\ref{fig:4gmass}~(left). All possible photon-pair assignments were considered, but generally only one assignment fell within the 2D $\eta\pi^0$ mass region. A clear peak corresponding to $\eta\pi^0\to 4\gamma$ events is seen. The mass resolution of the  $\pi^0$ and $\eta$ candidates are $\sigma_{\pi^0} = 7.6\;\text{MeV}$ and $\sigma_{\eta} = 19.1\;\text{MeV}$, respectively.  We select our signal to be in a rectangular region $\pm 3~\sigma_{\pi^0, \eta}$ around the $\eta\pi^0$ peak. The purity in this region is 81\%. We remove the background contribution using a two-dimensional sideband subtraction, as illustrated in Fig.~\ref{fig:4gmass}~(left), defining the $\pi^0$ sidebands to be 4.0~to~5.5~$\sigma_{\pi^0}$ away from the $\pi^0$ peak and the $\eta$ sidebands to be 4.0~to~6.0~$\sigma_{\eta}$ away from the $\eta$  peak.  Events in the $\pi^0$ and $\eta$ sidebands are weighted by a factor of $-2$ and $-3/2$, respectively, and the events in the corner regions were weighted by a factor of $+3$ to correct for over-subtraction. The $M(\eta\pi^0)$ mass distribution for the signal region after the sideband subtraction shows two clear peaks near the masses of the $a_0(980)$ and $a_2(1320)$ mesons (see Fig.~\ref{fig:4gmass}~(right)).

The backgrounds were studied using simulated samples of generic photoproduction events and various exclusive reactions.  These studies show that the background is primarily due to events with one charged particle and more than four photons, where the additional photons were not reconstructed.  For events with $M(\eta\pi^0)>1$~GeV, the largest background is due to $\gamma p \to b_1(1235) p$, with $b_1(1235) \to \omega \pi^0 \to (\gamma \pi^0) \pi^0 \to 5\gamma$. This background is concentrated between the $a_0(980)$ and $a_2(1320)$ peaks and is found to be removed efficiently by the sideband subtraction.

\begin{figure*}[!tb]
    \includegraphics[width=.85\textwidth]{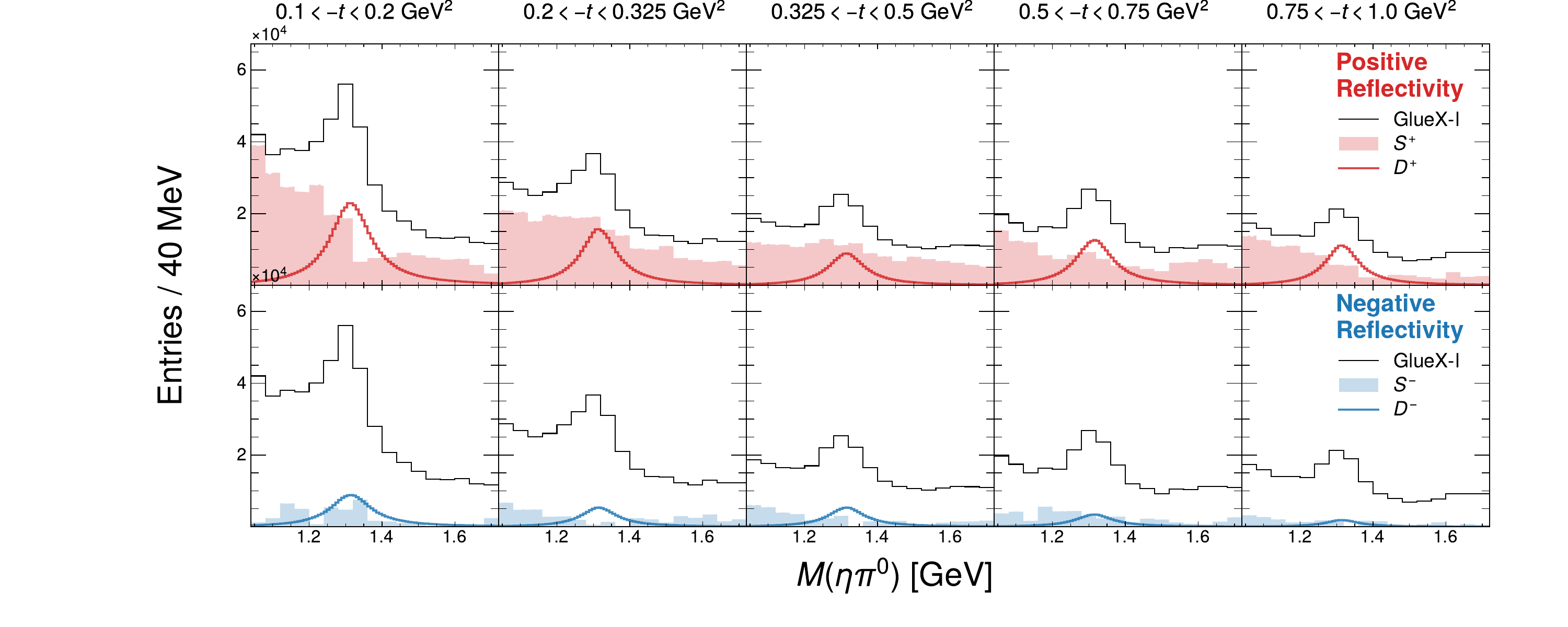}
    \caption{Intensities of the $S_0$-wave amplitudes (shaded histograms) and the coherent sums of the $D$-wave amplitudes (solid curves) with positive (top row) and negative reflectivity (bottom row) for individual $t$ bins (columns). The total measured intensity corrected for experimental acceptance is shown by the black histogram.}
    \label{fig:stamp}
\end{figure*}

\noindent\textit{Partial-Wave Analysis~---}  We perform a partial-wave analysis of the selected data in the $\eta\pi^0$ mass region $1.04 < M(\eta\pi^0) < 1.72\,\text{GeV}$, motivated by excluding the lower mass region dominated by the $a_0(980)$ signal and the high-mass region populated by baryon resonance production and other processes. In this approach, we decompose the data in terms of interfering partial-wave amplitudes that correspond to the production of intermediate states with well-defined spin-parity ($J^P$) quantum numbers decaying into $\eta\pi^0$. Since both daughter particles are spinless, the total angular momentum $J$ of the $\eta\pi^0$ system is identical to their relative orbital angular momentum $\ell$. In a given $t$ bin, the observed intensity distribution is described using the equation
\begin{widetext}
\begin{align}
 \nonumber    I(\Omega,\Phi,M)=2\kappa\left((1-P_{\gamma})\left[\sum_{\ell,m}[\ell]_{m}^{-}(M,\vec{x})\,\Re[Z_{\ell}^{m}(\Omega,\Phi)]\right]^2 \right. +(1-P_{\gamma})\left[\sum_{\ell,m}[\ell]_{m}^{+}(M,\vec{x})\,\Im[Z_{\ell}^{m}(\Omega,\Phi)]\right]^2 \\ 
   \left.+(1+P_{\gamma})\left[\sum_{\ell,m}[\ell]_{m}^{+}(M,\vec{x})\,\Re[Z_{\ell}^{m}(\Omega,\Phi)]\right]^2
    +(1+P_{\gamma})\left[\sum_{\ell,m}[\ell]_{m}^{-}(M,\vec{x})\,\Im[Z_{\ell}^{m}(\Omega,\Phi)]\right]^{2}\right),
    \label{eqn:zlm_intensity}
\end{align}
\end{widetext}
where $Z_{\ell}^{m}(\Omega,\Phi)=Y_{\ell}^{m}(\Omega)e^{-i\Phi}$ are phase-rotated spherical harmonics that describe the decay of the intermediate state into $\eta \pi^0$, $\Omega$ is the direction of the $\eta$ in the $\eta \pi^0$ helicity frame, $\Phi$ is the angle between the beam-polarization plane and the reaction plane defined by the beam, target, and recoil proton,  $P_{\gamma}$ is the fraction of linear polarization in the beam, $\ell$ is the orbital angular momentum between $\eta$ and $\pi^0$, $m$ is the associated spin-projection quantum number, and $\kappa$ is an overall phase-space factor.
The partial-wave amplitudes are given by 
\begin{equation}
    [\ell]_{m}^{\epsilon} = \sum_j \mathcal{C}^\epsilon_{\ell,m,j}\mathcal{D}^\epsilon_{\ell,m,j}(M, \vec{x}_j),
\end{equation}
where $\mathcal{C}$ represents the unknown coupling strengths and $\mathcal{D}$ parameterizes the dependence of the partial-wave amplitude on the invariant mass $M$ of the $\eta\pi^0$ system. $\mathcal{D}$ may also depend on a number of parameters $\vec{x_j}$, such as the mass and width of a resonance. 
We neglect possible incoherences arising from the nucleon spin and assume that the partial-wave amplitudes are fully coherent. Finally, the amplitudes are written in the reflectivity basis as introduced in Ref.~\cite{Mathieu:2019fts}. The reflectivity is indicated by the superscript $\epsilon=\pm$. In the high-energy limit and at low $-t$, the reflectivity corresponds to the naturality of the $t$-channel exchange.  

We include $S$- ($\ell=0$) and $D$-wave ($\ell=2$) contributions into the intensity model. In the reflectivity basis, the spin-projection quantum number of a tensor meson can take on five values, i.e. $-2 \leq m \leq +2$, so that the full waveset consists of the $\ell_{m}^{\epsilon} = S_0^\pm$ and $D^{\pm}_{-2, -1, 0, +1, +2}$ waves. We do not include $P$-wave or waves with $l > 2$ in our fits, as they are expected to be small~\cite{Afzal:2024ulu}.

We assume that the $D$-wave amplitudes are saturated by contributions from two tensor mesons, the $a_2(1320)$ and the much broader $a_2(1700)$, and model the corresponding $\mathcal{D}_{l, m, j}^\epsilon$ terms by relativistic Breit-Wigner amplitudes. The resonance parameters of the $a_2(1320)$ are initialized to the PDG average values~\cite{Workman:2022ynf} but are floating in the fits, using Gaussian constraints to include information on the uncertainty in these parameters.  
For the $a_2(1320)$ mass constraint, we use the uncertainty of the PDG average.  
The uncertainty of the PDG average for the $a_2(1320)$ width is smaller than the mass resolution, so we use the resolution in the constraint instead.
This yields constraints of $m_{a_2(1320)}=1318.2\pm0.6$~MeV and $\Gamma_{a_2(1320)}=111.1\pm5.5$ MeV. 
In contrast, due to the smaller contribution of the $a_2(1700)$, its mass and width parameters are fixed to the PDG average values: $m_{a_2(1700)}=1698$~MeV, $\Gamma_{a_2(1700)}=265$ MeV~\cite{Workman:2022ynf}.
Since we lack a realistic model for the $S$-wave amplitudes, we parameterize this contribution using piecewise-constant functions, effectively representing a \textit{binned}, model-independent approach without any assumption about the resonance content. For each 40 MeV wide $\eta \pi^0$ mass bin, we allow the magnitudes and phases of the $S$-wave amplitudes to float. To fix the unmeasurable global phase for each reflectivity, we fix the $S$-wave amplitudes to be real-valued and positive in one chosen mass bin.

\begin{figure}[!tb]
    \includegraphics[width=.48\textwidth]{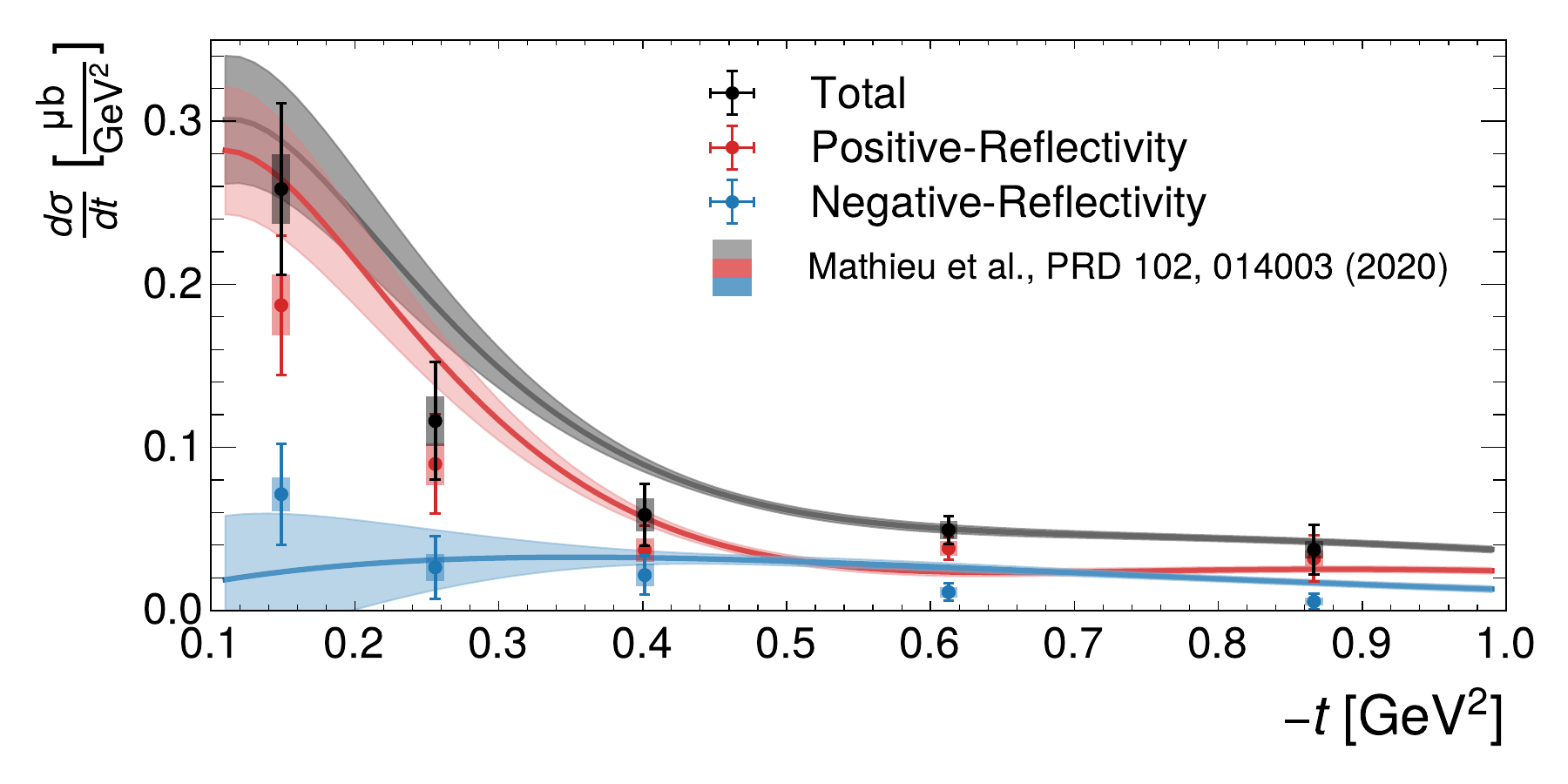}
    \caption{Differential cross section $d\sigma/dt$ for the reaction $\gamma p\to a_2(1320) p$ for $8.2 < E_\gamma < 8.8\,$GeV (black points). The positive- and negative-reflectivity components of the cross section are shown as red and blue points, respectively. Shaded boxes indicate statistical uncertainties as determined from bootstrapping the data sample, while the error bars indicate the total uncertainties including systematics. The curves show the prediction from the TMD model from Ref. \cite{Mathieu:2020zpm} using the same color code.}
    \label{fig:a2xsec}
\end{figure}

The parameters $\mathcal{C}_{\ell,m,j}^{\epsilon}$ and $\vec{x}_j$ are 
estimated by fitting the intensity model in Eq. \eqref{eqn:zlm_intensity} to the data using the unbinned extended maximum-likelihood approach, which takes into account the detector acceptance through simulated phase-space distributed Monte Carlo samples. The background contributions due to accidental beam photons (see Ref.~\cite{SuppMaterials}) and $\eta\pi^0$ sidebands are subtracted statistically at the level of the likelihood function using event weighting. The likelihood function is formulated and evaluated using the AmpTools package~\cite{amptools}. The optimization of the fit parameters is performed independently in each of the five $t$ bins using the \texttt{MINUIT} package~\cite{JAMES1975343}. In each $t$ bin, the data for all four beam-polarization orientations are fitted simultaneously.

To stabilize the fit, we reduce the parameter space by constructing the intensity such that all $m$ projections of either the $a_2(1320)$ or $a_2'(1700)$ are each produced by a set of either unnatural or natural parity Regge exchanges, which we assume to have no relative phase shift with respect to each other, within each naturality~\cite{Collins:1977jy}.  Therefore the phase of the set of $a_2(1320)$ (or $a_2'(1700)$) amplitudes for a given naturality comes from the $a_2$ Breit-Wigner propagator, which is independent of $m$, and an overall production phase that we assume to be common for all $m$.  
This leaves four relative-phase parameters in total.

\begin{figure}[tb!]
\centering
\includegraphics[width=0.45\textwidth]{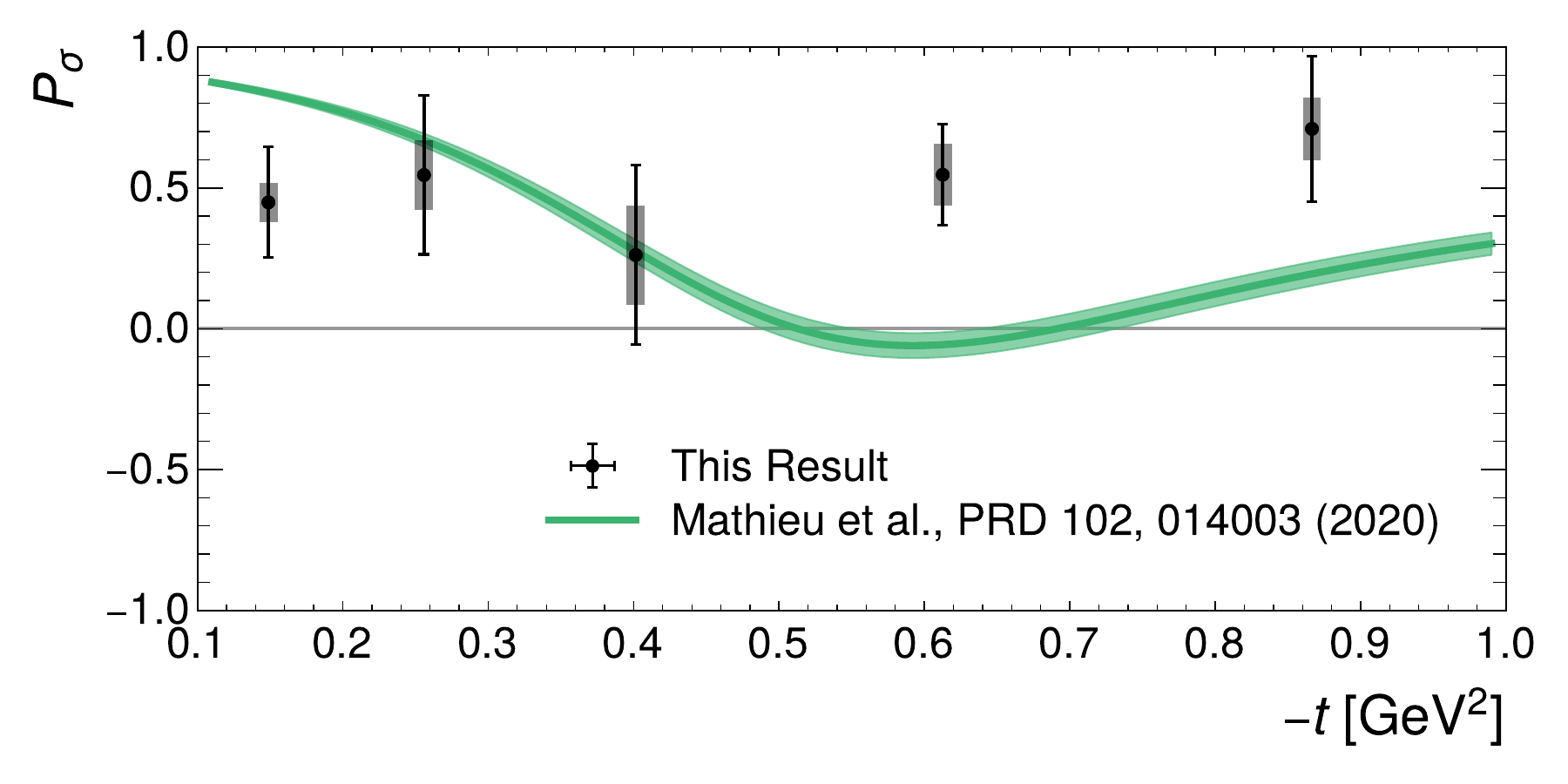}
\caption{The parity asymmetry defined in Eq.~(\ref{eqn:parity}) calculated from the values shown in Fig.~\ref{fig:a2xsec} (points with error bars). Values close to $+1/-1$ would indicate a dominance of positive/negative reflectivity production. The curve shows the prediction from the TMD model from Ref. \cite{Mathieu:2020zpm}.}
\label{fig:final_pa_result}
\end{figure}

Figure \ref{fig:stamp} shows the intensities of the $S_0$-waves and the coherent sums of the $D$-waves for $\epsilon = \pm$ in bins of $t$. The individual partial-wave cross sections are provided in Sec.\,V of the Supplemental Material~\cite{SuppMaterials}. 
Like the $S_0$-wave, we find that the $a_2(1320)$ is predominantly produced through natural-parity exchange. We observe a small $S_0$-wave contribution in the negative reflectivity throughout all $t$-bins, and a generally smooth larger contribution in the positive reflectivity.  The largest deviation from smoothness is observed in the $S_0^+$-wave around $M(\eta\pi^0)=1.3$\,GeV in the lowest $-t$ bin, and appears to be correlated with an increase in the $S_0^-$-wave.
By integrating the intensity of the coherent sum of the $a_2(1320)$ amplitudes, we obtain the efficiency-corrected $a_2(1320)$ yield for each reflectivity in each $t$ bin. Using the PDG values for the branching fractions of $a_2(1320)\to\eta\pi^0$, $\eta\to\gamma\gamma$, and $\pi^0\to\gamma\gamma$ and the total integrated luminosity of the GlueX-I data set in the coherent photon peak of 104\,$\mathrm{pb}^{-1}$, we calculate the differential photoproduction cross sections $d\sigma^{\pm}/dt$ for natural and unnatural-parity exchanges. 

Statistical uncertainties are estimated using the bootstrap approach~\cite{Efron:1979,hesterberg2015teachers}. Various studies are performed to estimate the systematic uncertainties, which are discussed in detail in Sec.\,III of the Supplemental Material~\cite{SuppMaterials}. The statistical and systematic uncertainties are added in quadrature to obtain the total uncertainty.

Figure~\ref{fig:a2xsec}~shows the differential cross sections $d\sigma^\pm/dt$ for positive and negative reflectivity together with their sum for the five $t$ bins. The points are located at the mean $-t$ value for each bin. 
The corresponding parity asymmetry 
\begin{equation}
P_\sigma = \frac{d\sigma^+ / dt - d\sigma^- / dt}{d\sigma^+ / dt + d\sigma^- / dt}
\label{eqn:parity}
\end{equation}
exhibits no $-t$ dependence and is consistent with a constant value of approximately $+0.5$, as shown in Fig.~\ref{fig:final_pa_result}.

\noindent\textit{Results and Conclusions~---}
Comparing our measurements for $d \sigma/dt$, summed over both reflectivity states, to the differential cross section measured by CLAS at lower photon energy~\cite{CLAS:2020rdz}, we find no evidence of the dip at $-t\approx0.5$~GeV$^2$ that is observed in the CLAS measurements.
We compare our cross sections to the calculation of Ref.~\cite{Mathieu:2020zpm}, which was performed with two limited wavesets, referred to as the Minimal and Tensor Meson Dominance (TMD) models, due to the lack of data to constrain the helicity couplings at the photon-Reggeon-tensor vertex. We do not consider the Minimal model here, since it does not describe the dip in the CLAS cross section data. Compared to the waveset used in this paper, the TMD model is based on a smaller waveset that does not include the $D_{-2}^\pm$, $D_{-1}^+$, and $D_{+2}^-$ waves and does not constrain the relative phases of the $D$-wave $m$ states.  The observed reduction in cross section and nonobservation of a dip in our measurements are in qualitative agreement with the predictions of the TMD model.  Both of these trends are consistent with the energy dependence of the leading $\rho$ and $\omega$ Regge exchanges.  However, our measured parity asymmetry suggests a roughly $t$-independent relative contribution from axial-vector $b_1$ and $h_1$ Regge exchanges.  This differs strongly from the TMD model, where the axial-vector exchange was assumed to be negligible at small $-t$ and to increase strongly with $-t$, in order to describe the cross section dip at CLAS energies.  Either this assumption is then wrong, or the energy dependence of the axial-vector exchanges is different than expected.

If we apply the waveset assumed by the TMD model to the partial-wave analysis of our data, we obtain similar cross sections to those reported above, but find that this model describes the angular distributions of our data increasingly poorly at larger $-t$.
This implies a more complicated coupling between the helicities of the beam photon and the produced tensor meson than is currently captured by the TMD model.
We observe the largest $a_2(1320)$ contribution in the $D_{+2}^+$-wave in the lowest $-t$ bin,
which decays quickly with $-t$.  This is roughly consistent with what is expected from the TMD model (see SEc.\,V of the Supplemental Material~\cite{SuppMaterials}), and has a curious similarity to the dominance of helicity-2 amplitudes observed by the Belle experiment in the process $\gamma\gamma \to \eta\pi^0$~\cite{Belle:2009xpa}.

The good qualitative description of our cross section measurements by the TMD model of Ref.~\cite{Mathieu:2020zpm} implies that the Regge amplitudes for the photoproduction of the $a_2^0(1320)$ and $\pi^0$~\cite{JointPhysicsAnalysisCenter:2017del,Mathieu:2018mjw} are similar, with a dominant vector exchange and substantial subdominant axial-vector exchanges.  These observations and the measured values for the $a_2(1320)$ photoproduction amplitudes in the various $D$-waves will help in developing improved models for isovector meson photoproduction and therefore in searching for excited isovector tensor mesons and the isovector hybrid meson $\pi_1(1600)$ in photoproduction. 
A comparison of the $a_2^0(1320)$ cross sections presented here to the planned measurement of $a_2^-(1320)$ production in $\gamma p \to \eta \pi^- \Delta^{++}$, where pion exchange is expected to dominate, will provide additional insight into the photoproduction of tensor mesons at GlueX energies.

We acknowledge productive discussions with V. Mathieu and A. Szczepaniak.
We would like to acknowledge the outstanding efforts of the staff of the Accelerator and the Physics Divisions at Jefferson Lab that made the experiment possible. This work was supported in part by the U.S. Department of Energy, the U.S. National Science Foundation, the German Research Foundation, Forschungszentrum J\"ulich GmbH, GSI Helmholtzzentrum f\"ur Schwerionenforschung GmbH, the Natural Sciences and Engineering Research Council of Canada, the Russian Foundation for Basic Research, the UK Science and Technology Facilities Council, the Chilean Comisi\'{o}n Nacional de Investigaci\'{o}n Cient\'{i}fica y Tecnol\'{o}gica, the National Natural Science Foundation of China and the China Scholarship Council. 
This research used resources of the National Energy Research Scientific Computing Center (NERSC), a U.S. Department of Energy Office of Science User Facility operated under Contract No. DE-AC02-05CH11231. This work used the Extreme Science and Engineering Discovery Environment (XSEDE), which is supported by National Science Foundation grant number ACI-1548562. Specifically, it used the Bridges system, which is supported by NSF award number ACI-1445606, at the Pittsburgh Supercomputing Center (PSC).
This research was done using services provided by the OSG Consortium~\cite{osg07,osg09,https://doi.org/10.21231/906p-4d78,https://doi.org/10.21231/0kvz-ve57}, which is supported by the National Science Foundation awards \#2030508 and \#1836650.
This material is based upon work supported by the U.S. Department of Energy, Office of Science, Office of Nuclear Physics under contract DE-AC05-06OR23177.

\bibliography{main}

\end{document}


\setcounter{linenumber}{1000}
\preprint{Draft}
\title{Supplemental Material}

\date{\today}

\maketitle

\tableofcontents

\section{Basic Event Selections}

 We select events which have exactly four photon candidates and one recoil proton candidate, and apply loose selection criteria for charged particles and calorimeter showers that are common to most GlueX analyses.  Photon candidates are calorimeter showers that have a minimum energy of 100~MeV and are not matched to the projection of any charged-particle track.  Photon candidates are also required to be within the fiducial range of $2.5^\circ < \theta < 10.3^\circ$ or $\theta > 11.9^\circ$, where $\theta$ is the polar angle between the photon momentum in the lab frame and the nominal beam axis. This avoids regions too close to the beam hole and where the two electromagnetic calorimeters overlap.  A recoil-proton candidate is defined as a track reconstructed in the drift chamber with a measured time-of-flight and drift-chamber ionization loosely consistent with that expected for a proton track. The track is required to come from the target and to have a minimum momentum of 300~MeV. 
Beam photons are required to be in the coherent-peak range of $8.2 < E_\gamma < 8.8$~GeV, where the photon polarization is maximal. The tagged beam photons and the $4\gamma p$ events measured in the GlueX detector are matched using timing information.  The large photon beam flux and detector resolution effects lead to ``accidental'' combinations that satisfy the event selections but include photons beyond the one that initiated the reaction. This background is estimated using matches that are between 2 and 4 beam bunches away from the ``prompt'' peak that contains our signal. These events are weighted by a factor $-1/6$ to subtract the background from the signal region.
We select fully reconstructed events by performing a four-constraint kinematic fit imposing energy and momentum conservation. The fitted four-vectors of the final-state particles are used throughout this paper. The kinematic fit greatly improves their resolution due to the well-measured beam-photon energies. Requiring the $p$-value of the fit to be greater than 0.01 efficiently selects fully reconstructed events.

\section{Target Excitation Rejection}

\begin{figure*}[tb!]
    \centering
    \includegraphics[width=\textwidth]{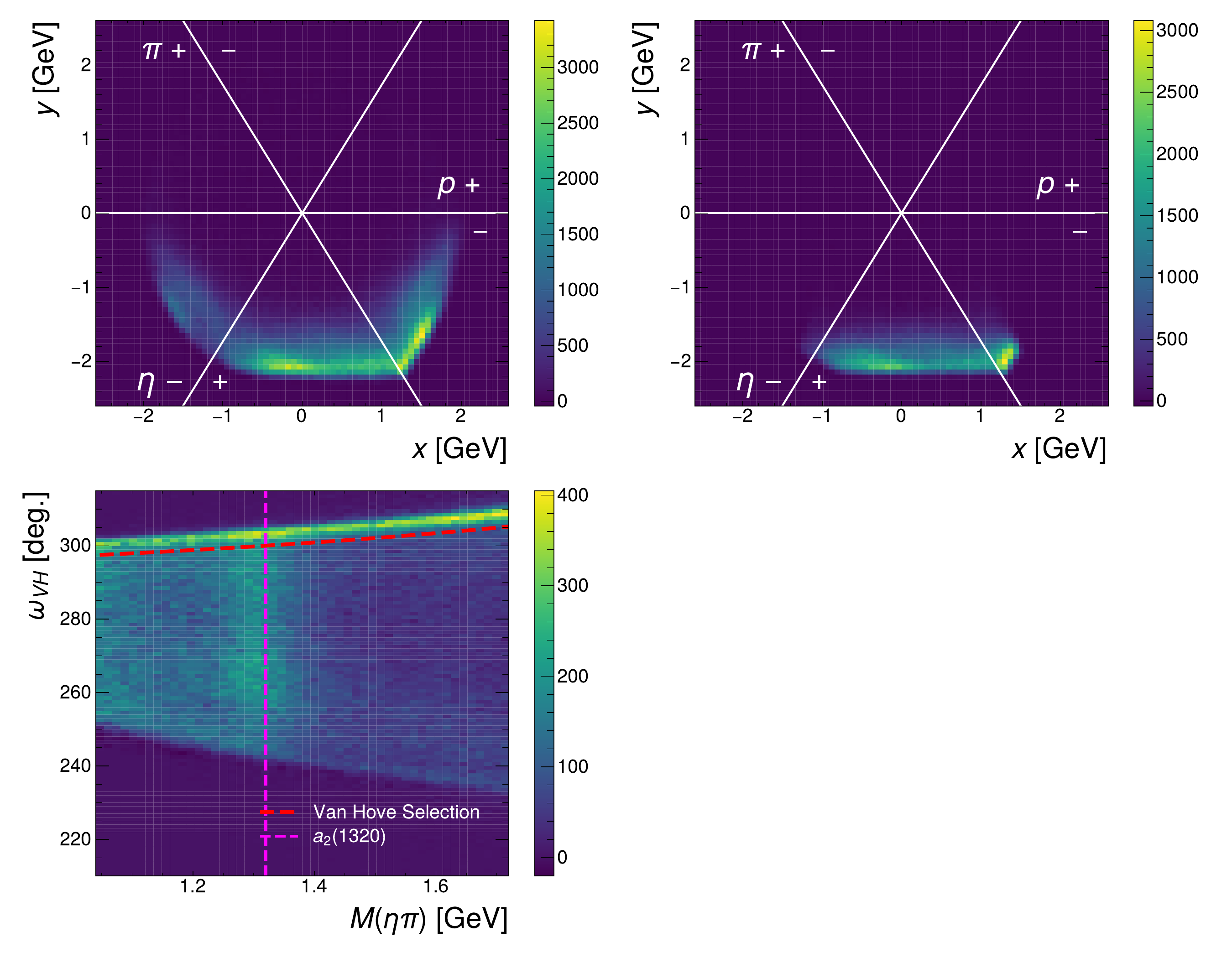}
    \caption{ Top row: Van Hove plots for the selected GlueX data with\ (right) and without\ (left) the additional selection on $1.04<M(\eta\pi^0)<1.72$ GeV. The $+/-$ signs designates regions where a particle is forward-/backward-going in longitudinal momentum space. The bottom plot shows the mass-dependent Van Hove selection to remove $\pi^0 p$ baryon resonances (red dashed curve).}
    \label{fig:vanhove_selection}
\end{figure*}

One background to the meson production process is excitation of the target proton which leads to the same final state, in particular $\gamma p \to \eta (N^*,\Delta^*) \to \eta (\pi^0p)$.  Such contributions are clearly seen in our data, with $\Delta^0(1232)$ contributing dominantly at low $-t$, and higher excited $N^*$ and $\Delta^*$ starting to contribute as $-t$ increases.  In order to efficiently reject such contributions over the wide analyzed $t$ range, a longitudinal phase space analysis was performed~\cite{VanHove:1969xa,Pauli:2018srg}.  In the center-of-mass (CM) frame of the reaction, $\Delta^*$ target excitation backgrounds are identified as having a forward-going $\eta$ and backward-going $\pi^0$ and $p$. A mass-dependent selection of $\omega_\text{VH} < 29^\circ \cdot \,\tan^{-1}[-1.05~\mathrm{GeV}^{-1} \cdot \,M(\eta\pi^0)+2.78]+328^\circ$ 
is found to reject these contributions. Here, $\omega_\text{VH}$ is the Van Hove angle that satisfies the relations $x=q\cdot \text{cos}(\omega_\text{VH})$ and $y=q\cdot \text{sin}(\omega_\text{VH})$, 
$M(\eta\pi^0)$ is the invariant mass of the $\eta\pi^0$ system, $q_i$ is the longitudinal momentum component of the $i$th final state particle with respect to the beam direction in the CM frame, and $q=\sqrt{q_1^2+q_2^2+q_3^2}$. The bottom panel of Fig.\,\ref{fig:vanhove_selection} illustrates the dependence of $\omega_\text{VH}$ on $M(\eta\pi^0)$, while the top panel illustrates the effect of the $M(\eta\pi^0)$ mass window applied for this analysis on the Van Hove plots.

\section{Systematic Uncertainties}
Contributions to the systematic uncertainty due to data selection criteria (missing mass, $\chi^2$ of the kinematic fit, unused energy, photon fiducial cuts, photon energy threshold, vertex cut, baryon rejection cut) are estimated by varying the respective criteria (Table \ref{tab:a2_systematics_study_rel}, Item 1). The degree and angle of the beam-photon polarization are fixed in the fit and their influence on the result is also evaluated by varying these values within their uncertainties (Table \ref{tab:a2_systematics_study_rel}, Item 2). Uncertainties associated with the partial wave analysis stem from the choice of the fit range in $M(\eta\pi^0)$ (Table \ref{tab:a2_systematics_study_rel}, Item 3), the fit model, and the number and choice of parameters to be optimized. The uncertainty due to the piecewise-constant parametrization of the $S_0$-wave amplitudes is determined by instead using 2nd-order polynomials to describe the $M(\eta\pi^0)$ dependence of their intensities and phases (Table \ref{tab:a2_systematics_study_rel}, Item 4). 
The resonance parameters of the $a_2(1320)$ are floating in the fit using Gaussian constraints. 
Larger and smaller widths of the Gaussian constraint on the $a_2(1320)$ parameters are used to estimate the corresponding uncertainty. The mass and width of the $a_2(1700)$ are fixed in our nominal fit, but are individually allowed to float in order to estimate the systematic uncertainty due to this assumption.  The systematic uncertainty obtained from the above studies is given in Table I, Item 5. 
We also take into account the systematic uncertainty from the beam-photon flux measurement (Table \ref{tab:a2_external_systematics}, Item 1) and from the recoil-proton reconstruction efficiency (Table \ref{tab:a2_external_systematics}, Item 2). Finally, we estimate the systematic uncertainty from the photon-reconstruction efficiency by studying the distribution of photons hitting the forward and barrel calorimeter, respectively, and extracting a weighted average of the detection systematic uncertainty (Table \ref{tab:a2_external_systematics}, Item 3). This second set of contributions to the systematic uncertainty is not dependent on the selection criteria and specifics of this analysis as well as the $t$ bin. Therefore, these uncertainties are called external systematics and summarized separately in Table \ref{tab:a2_external_systematics}. The total relative systematic uncertainty in each $t$ bin is taken to be the quadratic sum of all the items in Tables \ref{tab:a2_systematics_study_rel} and \ref{tab:a2_external_systematics}.

\begin{table*}[htb!]
    \caption{Systematic uncertainties relative to the nominal cross sections for each $t$ bin from low to high $-t$.} 
    \begin{tabular}{l|l|C{1.5cm}|R{1.5cm}|R{1.5cm}|R{1.5cm}|R{1.5cm}|R{1.5cm}}
    \hline
       Item & Systematic Study & $\epsilon$ & \multicolumn{5}{c}{$d\sigma^\epsilon/dt$ Relative Systematic Uncertainty [\%]} \\
    \hline \hline
    \multirow{2}{*}{1} & \multirow{2}{*}{Event Selections} & + & 3.0 & 24.1 & --- & 1.4 & 34.3 \\ 
    &  & -- & 3.3 & 36.1 & 19.1 & --- & 67.0 \\ \hline 
    \multirow{2}{*}{2} & \multirow{2}{*}{Beam Photon Polarization} & + & 0.2 & 1.0 & 0.6 & 0.7 & 1.0 \\ 
    &  & -- & 0.9 & 4.9 & 2.9 & 3.5 & 9.6 \\ \hline 
    \multirow{2}{*}{3} & \multirow{2}{*}{$M(\eta\pi^0)$ Fit Range} & + & 6.1 & 3.7 & 16.0 & 1.2 & 14.7 \\ 
    &  & -- & 6.2 & 17.5 & 6.7 & 13.1 & 43.3 \\ \hline 
    \multirow{2}{*}{4} & \multirow{2}{*}{$S_0$-Wave Parameterization} & + & 0.4 & 6.8 & 26.0 & 3.0 & 6.0 \\ 
    &  & -- & 35.8 & 5.9 & 37.3 & 32.9 & 3.3 \\ \hline 
    \multirow{2}{*}{5} & \multirow{2}{*}{Breit-Wigner Parameterization} & + & 14.6 & 12.1 & 12.3 & 2.5 & 13.7 \\ 
    &  & -- & 13.7 & 49.2 & 4.3 & 3.0 & 17.8 \\ \hline 
    \hline
    \end{tabular}
    \label{tab:a2_systematics_study_rel}
\end{table*}

\begin{table*}[htb!]
    \caption{External systematic uncertainties.}
    \begin{tabular}{l|l|R{5cm}}
    \hline
    Item & Source & Relative Systematic Uncertainty\\
    \hline
    \hline
    1 & Final State Photon Reconstruction & 11.4\% \\
    2 & Proton Reconstruction & 3\% \\
    3 & Flux Normalization & 5\% \\
    \hline
    \end{tabular}
    \label{tab:a2_external_systematics}
\end{table*}
\FloatBarrier
\clearpage
\section{Partial Wave Fit Quality} 

\begin{figure*}[htb!]
	\centering
	\includegraphics[width=\textwidth]{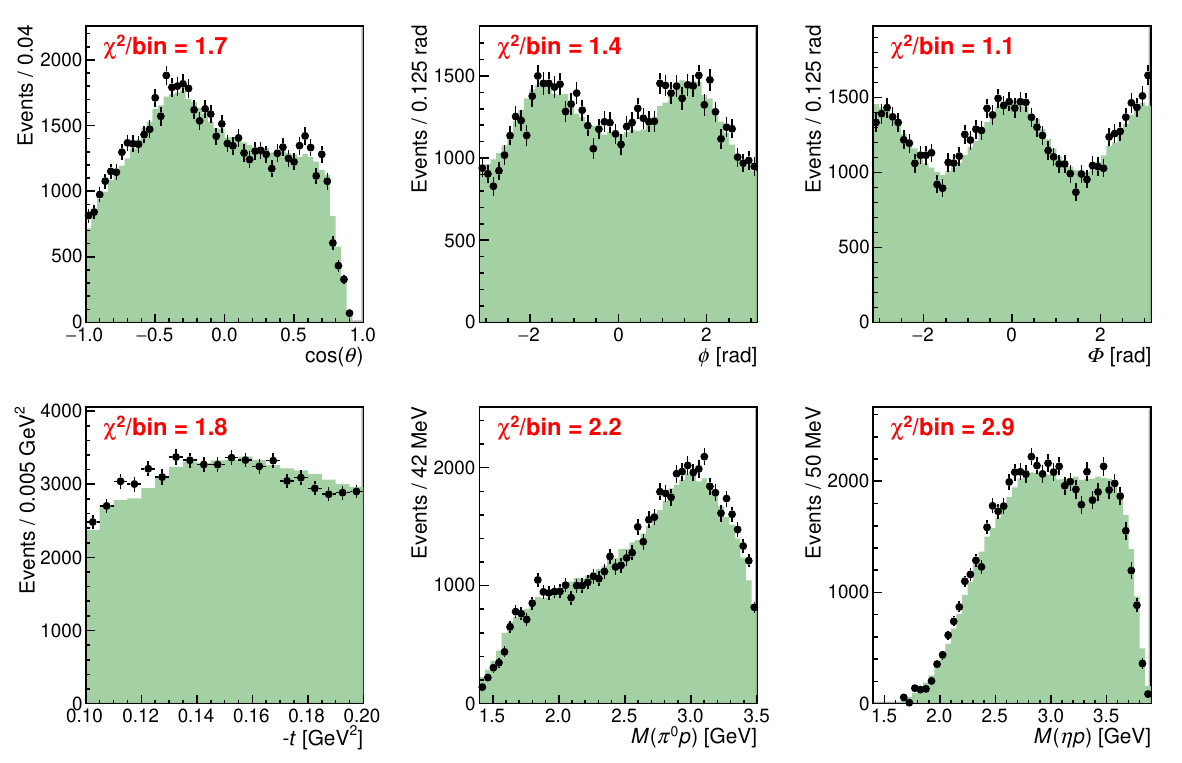}
	\caption{Fit quality in the bin $0.1<-t<0.2$ $\textrm{GeV}^2$. Black dots with error bars represent the observed GlueX data. Green shaded histograms show the reconstructed phase-space distributed Monte Carlo dataset weighted by the fit results. A $\chi^2$ per bin between these two distributions is shown in red in the corner.}
	\label{fig:fitQuality1}
\end{figure*}

To illustrate the quality of the partial wave fit, Fig.~\ref{fig:fitQuality1} overlays the fitted model on top of the GlueX data for the bin $0.1 < -t < 0.2$~GeV$^2$ as an example. As introduced in the paper, $\Omega=(\theta,\phi)$ are the angles of the $\eta$ in the $\eta\pi^0$ helicity frame and $\Phi$ is the angle between the reaction plane and the beam polarization plane. Backgrounds from the sidebands have been subtracted. For each one-dimensional projection of the kinematic distribution, a $\chi^2$ value per bin is provided in the respective subplot as a relative distance measure. From this value it can be seen that fit describes the data well.  For the other four $t$ bins we find a comparable level of agreement.

\section{Cross section and parity asymmetry values}

\begin{figure*}[htb!]
    \centering
    \includegraphics[width=\textwidth]{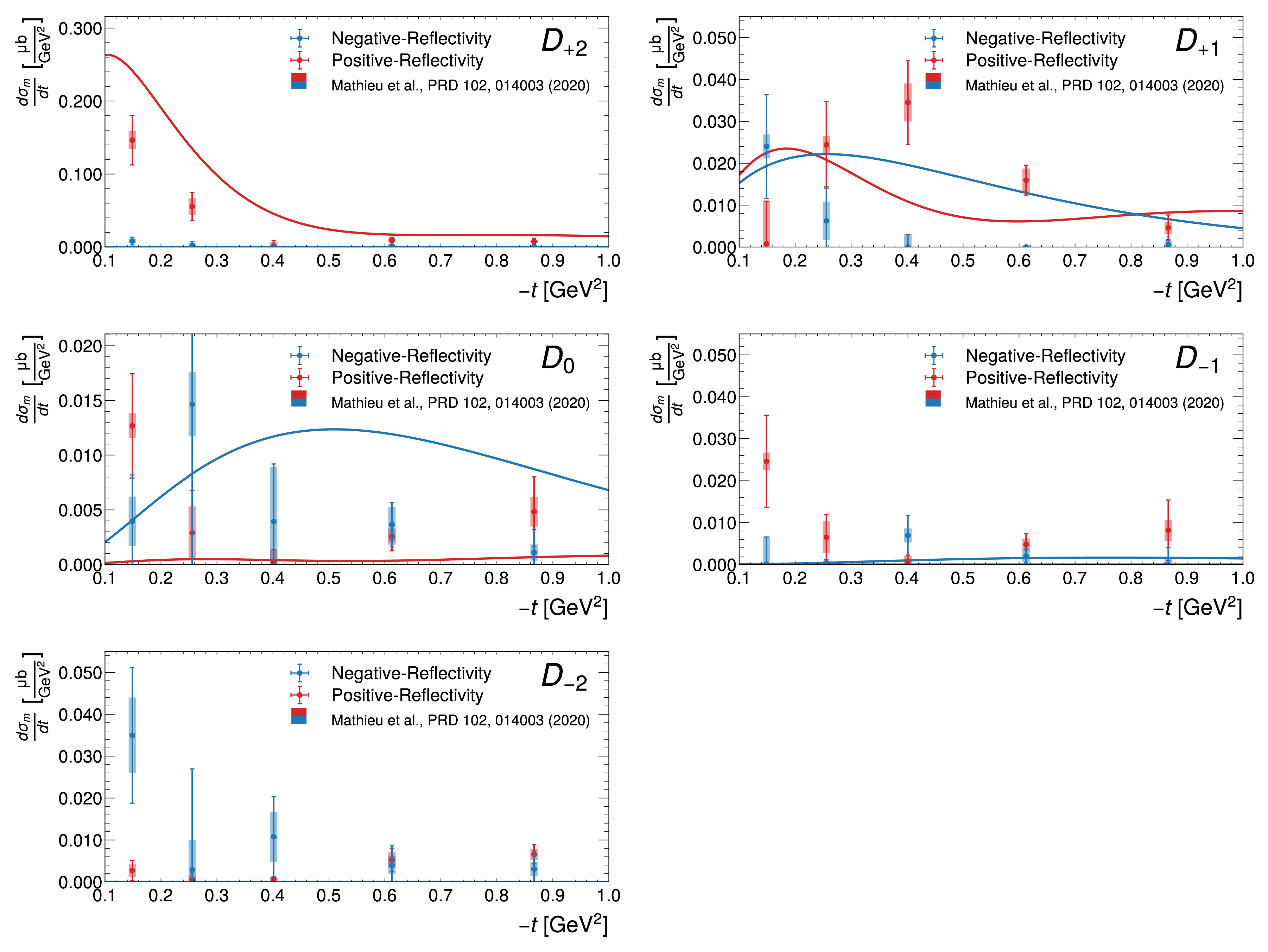}
    \caption{Differential $a_2(1320)$ cross sections $d\sigma_m/dt$ for positive (red) and negative (blue) reflectivity and for each spin-projection $m$. Shaded boxes indicate statistical uncertainties as determined from bootstrapping the data sample, while the error bars indicate the total uncertainties including systematics. The curves show the prediction from the TMD model from Ref.\cite{Mathieu:2020zpm} using the same color code.}
\label{fig:mproj_xsec}
\end{figure*}

In addition to the differential cross section shown in Fig.\,3 of the paper, the TMD model from Ref.\cite{Mathieu:2020zpm} also predicts the cross section for the $a_2(1320)$ produced with a particular spin projection and reflectivity. As mentioned in the paper, the TMD model assumes zero contribution from the waves $D_{-2}^{\pm}, D_{-1}^{+}, D_{+2}^{-}$. Figure \ref{fig:mproj_xsec} shows the measured differential cross sections for each spin projection $m$ across the five $t$ bins overlaid with the TMD model predictions. 

Despite the good qualitative agreement of the TMD model with the data for $-t \lesssim 0.5\;\text{GeV}^2$ (see Fig.~3 in the main paper), significant deviations are observed for the cross sections when split into the individual $m$-states. The largest observed contribution from $m = +2$ is still the largest contribution in the TMD model, but the model overpredicts this cross section, in particular in the bin $0.325 < -t < 0.5\;\text{GeV}^2$. The model predictions for $m^\epsilon = -1^-$, $0^-$, $+1^+$, and $+2^-$ are of similar magnitude as the measured values. However, the $a_2(1320)$ cross sections for $m^\epsilon = -1^+$, $-2^+$, and $-2^-$ that are zero in the TMD model are measured to have significantly non-zero values at least in some $t$ regions.
It is important to note that the TMD model was fitted to the CLAS $a_2(1320)$ differential cross section data only. The CLAS experiment did not perform a partial-wave analysis, and their photon beam was not polarized. Table \ref{tab:mproj_xsec} gives individual values including the associated statistical and systematic uncertainties.  Tables IV and V give the cross-section and parity-asymmetry values shown in Figs. 3 and 4, respectively, in the paper.

\begin{table*}[htb!]
    \caption{Differential $a_2(1320)$ cross sections $d\sigma_m^\epsilon/dt$ measured for each reflectivity $\epsilon$ and each spin projection $m$, as shown in Fig.~\ref{fig:mproj_xsec}. The first uncertainty is statistical and the second is systematic.}
    
    \begin{tabular}{l|l|l|l }
    \toprule
Wave & 
 \multicolumn{1}{c|}{ $-t$ range $[\mathrm{GeV}^2]$  }  
   &  \multicolumn{1}{c|}{$d\sigma^+_m / dt~~[\mathrm{nb} / \mathrm{GeV}^2]$}
   &  \multicolumn{1}{c}{$d\sigma^-_m / dt~~[\mathrm{nb} / \mathrm{GeV}^2]$} \\
\hline
$D_{-2}$ & $0.1<-t<0.2$ & $2.7\pm1.5\pm1.9$ & $35.0\pm9.0\pm13.4$ \\
 & $0.2<-t<0.325$ & $0.5\pm1.1\pm1.1$ & $2.9\pm7.1\pm23.0$ \\
 & $0.325<-t<0.5$ & $0.7\pm0.5\pm0.4$ & $10.8\pm6.0\pm7.5$ \\
 & $0.5<-t<0.75$ & $5.2\pm1.8\pm2.1$ & $4.0\pm2.1\pm4.1$ \\
 & $0.75<-t<1.0$ & $6.6\pm1.3\pm1.9$ & $3.0\pm1.7\pm3.9$ \\
\hline
$D_{-1}$ & $0.1<-t<0.2$ & $24.6\pm2.1\pm10.8$ & $0.1\pm6.4\pm0.7$ \\
 & $0.2<-t<0.325$ & $6.5\pm3.8\pm3.8$ & $0.3\pm0.6\pm0.7$ \\
 & $0.325<-t<0.5$ & $0.3\pm1.8\pm0.7$ & $7.0\pm1.7\pm4.5$ \\
 & $0.5<-t<0.75$ & $4.8\pm1.4\pm2.1$ & $2.1\pm1.6\pm0.5$ \\
 & $0.75<-t<1.0$ & $8.2\pm2.5\pm6.7$ & $0.0\pm1.5\pm3.7$ \\
\hline
$D_0$ & $0.1<-t<0.2$ & $12.7\pm1.1\pm4.6$ & $4.0\pm2.3\pm3.6$ \\
 & $0.2<-t<0.325$ & $2.9\pm2.4\pm3.1$ & $14.6\pm2.9\pm18.2$ \\
 & $0.325<-t<0.5$ & $0.2\pm1.2\pm0.1$ & $3.9\pm5.0\pm1.7$ \\
 & $0.5<-t<0.75$ & $2.6\pm0.7\pm1.1$ & $3.6\pm1.6\pm1.2$ \\
 & $0.75<-t<1.0$ & $4.8\pm1.3\pm2.9$ & $1.1\pm0.7\pm2.0$ \\
\hline
$D_{+1}$ & $0.1<-t<0.2$ & $0.8\pm10.1\pm0.9$ & $24.0\pm2.9\pm12.1$ \\
 & $0.2<-t<0.325$ & $24.4\pm2.1\pm10.1$ & $6.2\pm4.6\pm6.7$ \\
 & $0.325<-t<0.5$ & $34.5\pm4.5\pm9.0$ & $0.0\pm3.1\pm0.0$ \\
 & $0.5<-t<0.75$ & $16.0\pm2.8\pm2.3$ & $0.0\pm0.4\pm0.1$ \\
 & $0.75<-t<1.0$ & $4.6\pm1.5\pm2.7$ & $0.6\pm0.8\pm0.8$ \\
\hline
$D_{+2}$ & $0.1<-t<0.2$ & $146.4\pm12.2\pm31.7$ & $8.2\pm2.3\pm4.7$ \\
 & $0.2<-t<0.325$ & $55.5\pm11.3\pm15.4$ & $2.3\pm2.8\pm4.1$ \\
 & $0.325<-t<0.5$ & $1.4\pm5.4\pm4.6$ & $0.0\pm2.7\pm0.0$ \\
 & $0.5<-t<0.75$ & $9.6\pm2.9\pm1.6$ & $1.5\pm2.4\pm0.6$ \\
 & $0.75<-t<1.0$ & $7.6\pm1.6\pm4.2$ & $0.7\pm0.9\pm3.0$ \\

\hline \hline
    \bottomrule
    \end{tabular}

    \label{tab:mproj_xsec}
\end{table*}

\begin{table*}[h!]
\caption{Differential $a_2(1320)$ cross sections $d\sigma^\epsilon/dt$ measured for reflectivity $\epsilon=\pm$ and their sum as shown in Fig.\,3 of the paper. The first uncertainty is statistical and the second is systematic.}

\centering
\begin{tabular}{l|l|l|l}
\hline\hline    

 \multicolumn{1}{c|}{ $-t$ range $[\mathrm{GeV}^2]$  }  
   &  \multicolumn{1}{c|}{$d\sigma^+ / dt~~[\mathrm{nb} / \mathrm{GeV}^2]$}
   &  \multicolumn{1}{c|}{$d\sigma^- / dt~~[\mathrm{nb} / \mathrm{GeV}^2]$} 
   &  \multicolumn{1}{c}{$d\sigma / dt~~[\mathrm{nb} / \mathrm{GeV}^2]$} \\

\hline
$0.1<-t<0.2$ \ & \ $187\pm19\pm39$ \ & \ $71\pm10\pm29$ \ & \ $259\pm21\pm48$ \\
$0.2<-t<0.325$ \ & \ $90\pm13\pm28$ \ & \ $26\pm8\pm17$ \ & \ $116\pm15\pm33$ \\
$0.325<-t<0.5$ \ & \ $37\pm7\pm13$ \ & \ $22\pm7\pm10$ \ & \ $59\pm10\pm16$ \\
$0.5<-t<0.75$ \ & \ $38\pm5\pm5$ \ & \ $11\pm3\pm4$ \ & \ $49\pm6\pm7$ \\
$0.75<-t<1.0$ \ & \ $32\pm5\pm13$ \ & \ $5\pm2\pm5$ \ & \ $37\pm5\pm14$ \\
\hline\hline
\end{tabular}

\label{tab:a2_xsec_final_results}
\end{table*}

\begin{table*}[h!]
\caption{Parity asymmetry for the reaction $\gamma p\to a_2(1320)p$ in $t$ bins as shown in Fig.\,4 of the paper. The first uncertainty is statistical and the second is systematic.}

\centering
\begin{tabular}{l|l}
\hline\hline

 \multicolumn{1}{c|}{ $-t$ range $[\mathrm{GeV}^2]$  } & \multicolumn{1}{c}{ $P_{\sigma}$ } \\
\hline

$0.1<-t<0.2$ \ & \ $0.45\pm0.07\pm0.18$ \\ 
$0.2<-t<0.325$ \ & \ $0.55\pm0.12\pm0.25$ \\ 
$0.325<-t<0.5$ \ & \ $0.26\pm0.18\pm0.27$ \\ 
$0.5<-t<0.75$ \ & \ $0.55\pm0.11\pm0.14$ \\ 
$0.75<-t<1.0$ \ & \ $0.71\pm0.11\pm0.23$ \\ 
\hline\hline
\end{tabular}
\label{tab:a2_pa_final_results}
\end{table*}

\bibliography{supplemental}